\documentclass[12pt,preprint]{aastex}

\slugcomment{Draft Version 2012 August 30}

\shorttitle{Magnetic Tilt Angles}
\shortauthors{Li, Ulrich}

\begin{document}

%% one line. However, you may use \\ to force a line break if you desire.

\title{Long-Term Measurements of Sunspot Magnetic Tilt Angles}

\author{Jing Li\altaffilmark{1} and Roger K. Ulrich\altaffilmark{2}}
\affil{$^1$Department of Earth and Space Sciences, University of California at Los Angeles, \\
Los Angeles, CA  90095-1567}
\affil{$^2$Department of Physics and Astronomy, University of California at Los Angeles, \\
Los Angeles, CA  90095-1567}
\email{jli@igpp.ucla.edu}

\begin{abstract}
Tilt angles of close to 30,600 sunspots are determined using Mount Wilson daily averaged magnetograms taken from 1974 to 2012, and MDI/SoHO magnetograms taken from 1996 to 2010. Within a cycle, more than 90\% of sunspots have a normal polarity alignment along the east-west direction following Hale's law. The median tilts increase with increasing latitude (Joy's law) at a rate of $\sim0^\circ.5$ per degree of latitude. Tilt angles of spots appear largely invariant with respect to time at a given latitude, but they decrease by $\sim 0^\circ.9$ per year on average, a trend which largely reflects Joy's law following the butterfly diagram. We find an asymmetry between the hemispheres in the mean tilt angles.  On average, the tilts are greater in the southern than in the northern hemisphere for all latitude zones, and the differences increase with increasing latitude. 
%A wide variance of sunspot tilt angles is measured by the Root-Mean-Square ({\it r.m.s.}). It is $\sim 27^\circ$ with the current measurements. 
%In the low, intermediate and high latitudinal zones, the median tilt angles were greater in the southern hemisphere than those in the northern hemisphere by $\sim5^\circ$, $\sim10^\circ$, and $\sim15^\circ$, respectively, during the time of study. 
%The unusual behavior of the current cycle is demonstrated in the sunspot tilt angles. By the end of the cycle 23, tilt angles had unusually large variations during the prolonged solar minimum. 
%The total magnetic flux varies with the magnetic area in average $380\pm 40$ [Mx cm$^{-2}$]. The median tilt angles over 300 days as function of time are clearly in sync with the solar cycles. During the solar minima of cycles 21 and 22, the tilt angles were close to zero. 
\end{abstract}

% The different journals have different requirements for keywords.  The
% keywords.apj file, found on aas.org in the pubs/aastex-misc directory, 
% contains a list of keywords used with the ApJ and Letters.  These are 
% usually assigned by the editor, but authors may include them in their 
% manuscripts if they wish. 


\section{Introduction}
The sunspot magnetic tilt angle is the angle between the east-west direction and the line connecting opposing polarity portions in a sunspot group or in a pair of sunspots. It can be described by two well-known but qualitative laws: Hale's and Joy's laws. Hale's law states that, at any moment, sunspot polarity pairs have opposite sign orientation between the northern and southern hemispheres. At any given cycle, a majority of the leading spots have opposite sign polarities in opposite hemispheres. The sign orientations switch at the onset of a new cycle when new sunspots emerge from high latitudes \citep{1925ApJ....62..270H}. Joy's law describes the latitude dependence of the tilt of bipolar sunspot regions: the leading spots are closer to the equator than the following spots and this tilt angle increases with increasing latitude, as found in a study by Joy reported by \citet{1919ApJ....49..153H}.  One motivation of the present study is to obtain quantitative and accurate, long-term measurements of the tilt angle as functions of time and latitude.

Accurate, long-term measurements of the tilt angle can be used to discriminate between published models. In the Babcock-Leighton model \citep{1961ApJ...133..572B,1969ApJ...156....1L}, which describes the repetitive solar global magnetic field patterns, the tilt angles of bipolar magnetic regions are interpreted as pitch angles of  magnetic field lines stretched by differential rotation. This interpretation by itself does not reproduce Joy's law. On the other hand, stretching magnetic field by differential rotation in the latitudinal and radial directions is the basis of the magnetic buoyancy needed for flux tube emergence \citep{1955ApJ...121..491P,1955ApJ...122..293P}. Numerical simulations show that magnetic buoyancy and the Coriolis force together generate the correct latitudinal tilt angle distribution \citep{1993A&A...272..621D,1994ApJ...436..907F,1995ApJ...438..463F}.  These models also predict a close correlation between the magnetic field strength and the tilt angle.  While some observations confirm the existence of such a correlation \citep{2003SoPh..215..281T,2010A&A...518A...7D},  others show no evidence for it.  For example, a study of the tilt angles of over 700 sunspot groups during their disk passage found that the sunspot tilt angles are established at an early stage of the sunspot emergence and remain roughly constant until the decay of the active regions \citep{2008ApJ...688L.115K}. A recent numerical simulation of a thin flux tube in a rotating turbulent spherical shell needs fewer constraints on the initial magnetic strengths and shows a statistical trend of the tilt angles with latitude \citep{2011ApJ...741...11W}. An initial flux strength near 40 kG produces tilt angles following Joy's law, while stronger and weaker flux tube strengths give a wider spread of tilt angles on the surface of the Sun. The differences between the Babcock-Leighton model and the magnetic flux buoyancy process underline the complex nature of the magnetic flux emergence as part of the solar dynamo operation. 

We measured the tilts of sunspots that erupted over the course of 38 years (1974-2012) using magnetic field data. Our analysis differs from an earlier one by \citet{1994SoPh..149...23H} which was based on the white light images from Mt. Wilson from 1917 to 1985 from which sunspot positions were measured by \citet{1984ApJ...283..373H}. In their study and others based on it, magnetic polarities were not known and the position of the polarity reversal was assumed to be midway between the leading and trailing spots. The second difference between our work and earlier studies is that we examined the average tilt angles of a sunspot group using one magnetogram per day, while others studied daily sunspot tilt angle variations \citep{1986ApJ...303..480G,1991SoPh..132...49H,1991SoPh..132..257H,2002SoPh..209..321M}. The time period that we covered was exactly following that period covered by \citet{2010A&A...518A...7D} for the tilt angle study. The various studies are also distinguished by their relative sizes: ours is larger than those studied by  \citet{1989SoPh..124...81W} and \citet{2008ApJ...688L.115K} because our time period is longer, but smaller than the sample in a recent study by \citet{2012ApJ...745..129S}. Their sample ``spans a range of scale sizes across nearly four orders of magnitude'' including sunspots. It is interesting that they found large-scale bipolar regions having tilt angles about $90^\circ$ from the average Joy's law distribution. These regions are larger than sunspot regions, are not included in our study.
%The tilt angles of these regions are probably associated with the polar field polarity change, but t

What mostly separates our study from many others using magnetograms is that we analyzed only identified sunspot and sunspot groups. By directly overlapping positions of sunspot groups on the magnetograms, we avoided the need to develop a more sophisticated algorithm for pattern recognition. We constructed magnetic bipolar regions by calculating centers of mass of the positive and negative polarities within a sunspot group. This approach results in the averaged tilt angles of a sunspot region, but not tilt angles due to local disturbances and fine magnetic structures. This is different from the tilt angles obtained over small-scale, quiet-Sun bipoles, or ephemeral regions comparable to the X-ray bright points \citep{2002ApJ...564.1042S,2010ApJ...717..357T}.  

\section{Data}
The data used in this study consist of sunspot daily records and longitudinal magnetograms.
For the period 1974 to 1990, we use the sunspot records made at MWO, which has assigned its own sunspot group numbers since 1920. The Mt. Wilson sunspot records do not contain the areas of the sunspot groups, but have the crucial information needed for this study, namely, the sunspot central meridian crossing time and latitude. From 1991 onward, the sunspot reports were prepared by Space Weather Forecast Center (SWFC) from observations made by MWO and other observatories. They are made available via their web site. Parameters relevant to this study include the observing dates, sunspot NOAA numbers, sunspot group areas, and sunspot group disk locations. They correspond to the sunspot status valid at 00:00 UT each day. 
%These official sunspot reports are not available prior to 1991. 

The magnetograms for the entire study period were obtained at the 150 foot solar tower telescope at Mount Wilson Observatory. A two-mirror coelostat is installed on top of the tower, reflecting light to the objective lens with a focal length of 150 feet. Solar images are formed at ground level in an observing room. A spectrograph is situated under the ground level in a vertical pit. Sunlight enters the spectrograph through a slit, and exits into one of several fiber-optic bundles at selected wavelengths. The first Mt. Wilson magnetograph was built by \cite{1953ApJ...118..387B} and daily magnetograms of the full disk of the sun were started in 1957. The magnetograph was largely rebuilt by \citet{1983SoPh...87..195H}. It consists of an interference filter having a 100 \AA~bandwidth centered on the Fe I $\lambda$5250 line, a KD*P crystal-Glan Thompson prism combination providing circular polarization modulation; a Littrow spectrograph with a new grating;  and a new exit slit assembly. The current  system was an upgrade to a 24-channel spectrophotometer by \citet{1991SoPh..135..211U,2002ApJS..139..259U} adding three spectral lines: Cr II $\lambda$5237, NaD $\lambda$5896 and NiI $\lambda$6768. Since the mid-1980s, daily observations at MWO generate the magnetograms and intensity-grams taken in the 5250.2 \AA~Fe I and 5237.3 \AA~Cr II lines. These data are used to model the total solar irradiance variations \citep{2010SoPh..261...11U}. 

The systematic upgrade of the magnetograph at MWO, while largely preserving the spectral sampling characteristics, provides consistent magnetic field observations over many decades. The full-disk daily-averaged magnetograms at the spectral line $\lambda$5250.2 \AA~have the longest observing history, and are used in this study. A daily magnetogram was constructed by averaging as many as 20 images corrected to the observing time at 20:00 UT. Differential rotation is taken into account when the images are combined. The non-uniform pixel size due to the nonuniform image sampling during the solar image scan was corrected. The final image size is $512\times512$ pixels, and the pixel size is $3.7\arcsec\times3.7\arcsec$. The daily averaged magnetograms are available from 1985 to the present time. Prior to 1985, the individual magnetograms obtained once a day were used for the study.

From 1996 January  to 2010 December, we employed the full disk MDI/SoHO  \citep{1995SoPh..162..129S} magnetograms, level 1.8, with a 96-minute time cadence. The MDI employs a Michelson interferometer as the tunable spectral device, centered at Ni I 6768 \AA. The image size is $1024\times1024$ pixels, and the pixel size is $\sim 2\arcsec\times2\arcsec$. A single MDI magnetogram per day was employed for all sunspots appearing on that day for the tilt angle calculations. That magnetogram was taken closest to 00:00 UT when the sunspots were recorded.

%The MDI data serve as an independent source to verify the tilt angle measurements obtained at MWO. After comparing the separate series,  we find that they are sufficiently similar that we can combine the tilt angle measurements of MWO and MDI into a single series. 

\section{Method}
\begin{figure}[t]
%\epsscale{1.2}
\begin{center}
%\plotone{tilt_def.png}
%\includegraphics[width=0.7\textwidth]{tilt_def.png}
\includegraphics[width=0.7\textwidth]{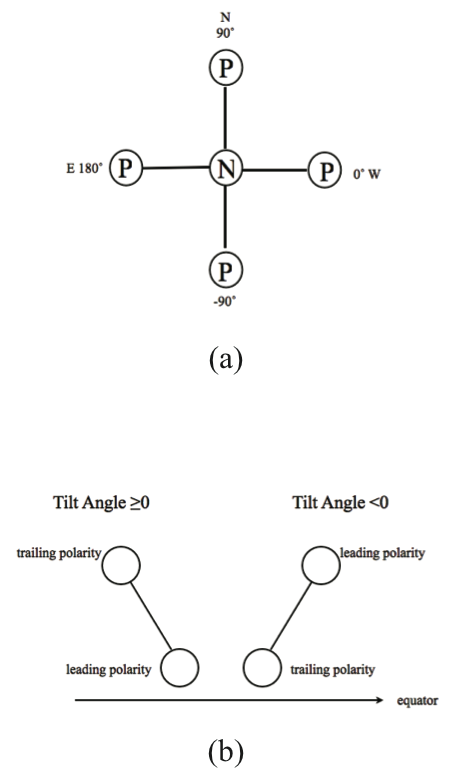}
\caption{Definitions of the tilt angles. (a) Tilt angles vary between  [$-180^\circ,180^\circ$]. ``P'' represents the positive polarity, and ``N'' represents the negative polarity. (b) Tilt angles vary between [$-90^\circ$, $90^\circ$]. The orientations of the tilt angles are identical in both definitions. The difference is that the leading spot polarity is indicated in the definition (a), but not in the (b). \label{tilt_def}}
\end{center} 
\end{figure}

Two definitions representing different aspects of the tilt angle are useful for studying Hale's and Joy's laws.  They  are illustrated in Fig. (\ref{tilt_def}).  Fig. (\ref{tilt_def}a) is similar to that defined by \citet{1991SoPh..132..257H}. The tilt angles, $\gamma$, vary from $\leq 90^\circ$ to $>90^\circ$ as the positive polarity component changes from the leading to the trailing component. This definition carries the information of the leading spot polarity signs. {\it Hale's} law is conveniently illustrated using this definition. The orientation of tilt angles in Fig. (\ref{tilt_def}b) is identical to those defined in Fig. (\ref{tilt_def}a), but the bipolar magnetic polarities are no longer reflected by the values of tilt angles. The tilt angles vary between [$-90^\circ,90^\circ$] \citep{1991ApJ...375..761W}, and are the most suitable to illustrate {\it Joy's} law. It is noted that signs of the tilt angles are the opposite of the conventional angles in a Cartesian coordinate system in Fig. (\ref{tilt_def}b). This ensures that Joy's law will be conveniently described as the tilt angles increasing with latitude.

The tilt angle measurements are performed by an automated IDL package which accomplishes the following tasks: (a) locate the sunspots on the magnetograms; (b) determine the centroids of the entire sunspot, positive and negative polarities within each spot; and (c) calculate the tilt angles for each sunspot group. Once the sunspot locations and sizes were initially determined,  the automated IDL program runs an iteration of tasks (b) and (c) until the tilt angle converges or stabilizes.

Specifically, we first employed an ellipse to define the boundary of an active region. The center of an elliptical area is the sunspot location tabulated by MWO and SWFC. The disk locations of the sunspots are corrected for differential rotation \citep{1951MNRAS.111..413N} in order to match the sunspots with the magnetograms. The orientation of the ellipse representing the tilt angle is set to $0^\circ$ because most sunspot groups are elongated in the E-W direction. After some experiments,  we found that reasonable tilt angles were produced if the long and short semi-axes, $a$ and $b$, of the  initial ellipse were set as

\begin{equation}
a=20\times (sunspot~ area)^{1/3}, ~~~~ b=a/2
\label{ss_size}
\end{equation}

\noindent where {\it sunspot area} is the area given by SWFC in millionths of the solar hemisphere, but was converted to image pixels. For sunspots prior to 1991, {\it sunspot area} is set to 60 millionth hemisphere. We note that  the value of the initial area does not appear critical for the tilt angle calculations as long as the sunspot locations are roughly correct. At least, this is true to the MWO magnetograms. Empirically, the search program converges to the region surrounding the sunspot groups. 
%We checked that the final bipolar orientation is consistent with that of the sunspot when viewed by eye. 

\begin{figure}[t]
%\epsscale{1.2}
\begin{center}
%\plotone{tilt_def.png}
%\includegraphics[width=0.6\textwidth]{ellipses.png}
\includegraphics[width=0.6\textwidth]{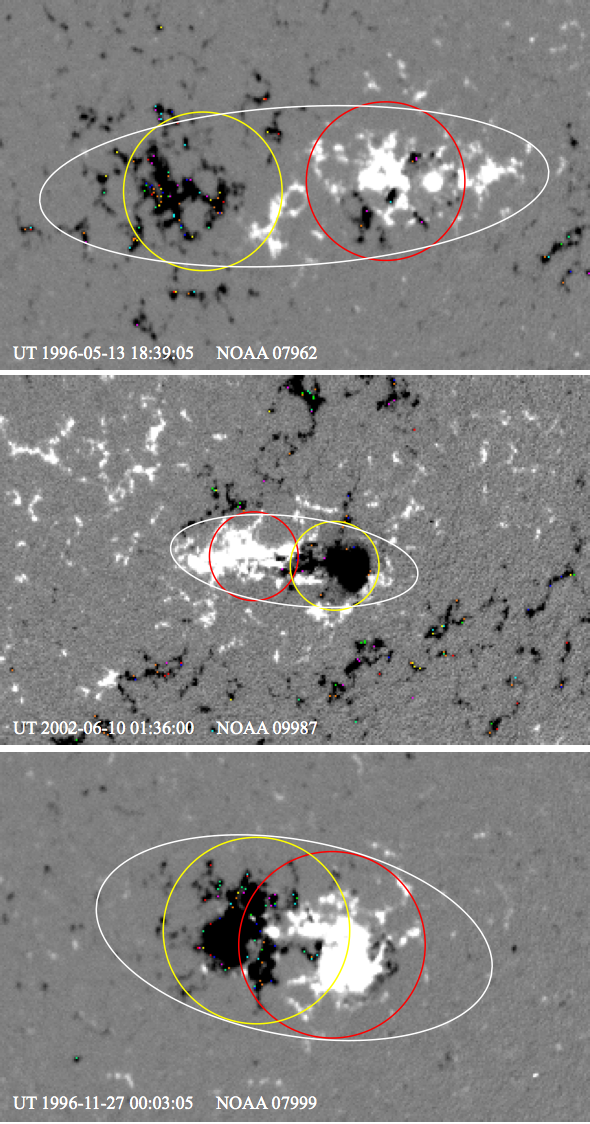}
\caption{The ellipses represent the sunspot group boundaries. The bipolar  spot pair is represented by red and yellow circles for positive and negative polarities. Three extreme cases are illustrated with polarity pair circles separate, contacted and overlapping. The long axes of the ellipses with respect to the equator represent the magnetic tilt angles of each bipole. Magnetic field strength is displayed on a scale from -250 (black) to 250 G (white). The size of each panel is 320$\times$200 pixels ($635\arcsec\times398\arcsec$). East is to the left, and the north is up. The magnetograms are from MDI data.\label{ellipses}}
\end{center} 
\end{figure}

Following the initial setup of the ellipse over a sunspot group, a new centroid of the active region is computed with the mass center of the total magnetic flux and the pixels are sorted according to their polarity. From these two pixel sets, we find a pair of circles defining a simple bipolar configuration, which radii are defined as $\sqrt{area/\pi}$, where $area$ is the total number of pixels in each polarity. The idealized bipolar regions often have overlapping circles of opposite polarity because pixels of either strong or weak magnetic fields are given equal counts to sum up the areas. In reality, the strong opposite polarities appear side by side in an active region, but weaker field spreads throughout the region. In our analysis, magnetic field signals are extracted down to the minimum field strength of 10 G for MWO data, and 20 G for MDI data. Figure (\ref{ellipses}) shows examples of three cases of separated, contacted and overlapping magnetic polarity regions fitted by our algorithm.

The measured centroids of the positive and negative polarities are in the observed solar disk coordinate system designated as [$x\arcsec,y\arcsec$]. We convert them into heliographic latitude and longitude, [$B,L$] degrees, taking into account the instantaneous heliographic latitude, $B_0$. The conversion is available with the code {\it xy2lonlat} which is implemented in the SolarSoftware \citep{1998SoPh..182..497F} by Thomas Metcalf. An updated magnetic tilt angle is calculated from $\gamma=\arctan(\Delta B,\Delta L\times \cos B)$, where $\Delta B$ and $\Delta L$ are the differences between positive and negative polarity centroids in heliographic latitude and longitude; $B$ is the  heliographic latitude of the positive polarity centroid.
% with the bounding ellipse and the two polarity circles marked.

\begin{figure}[t]
%\epsscale{1.2}
\begin{center}
%\plotone{tilt_def.png}
%\includegraphics[width=1.0\textwidth]{tilt_iteration.png}
\includegraphics[width=1.0\textwidth]{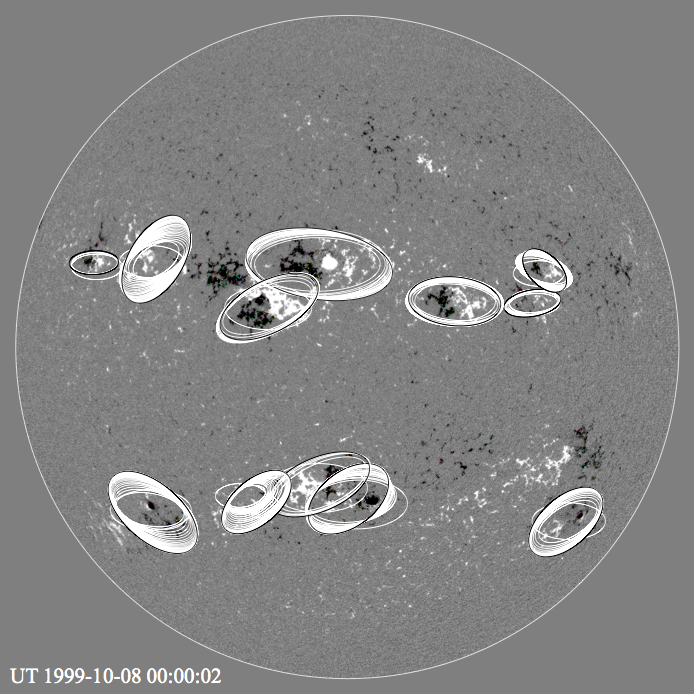}
\caption{Illustration of the tilt angle measurement algorithm. The multiple ellipses around each sunspot represent successive iterations of the fitting algorithm. The initial ellipses are parallel to the equator. They are quickly adjusted to the orientation close to the final state, shown as black ellipses. \label{tilt_it}}
\end{center} 
\end{figure}

\begin{figure}[t]
%\epsscale{1.2}
\begin{center}
%\plotone{tilt_def.png}
%\includegraphics[width=1.0\textwidth]{tilt_fin.png}
\includegraphics[width=1.0\textwidth]{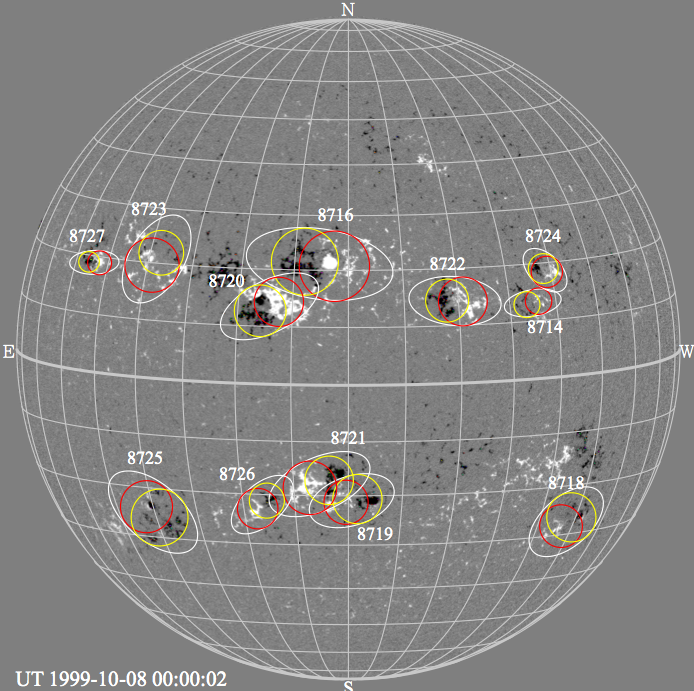}
\caption{Measured tilt angles are represented by the tilts of the long axes of ellipses with respect to the equator. Two circles inside the ellipses represent the mass centers and sizes of positive (red) and negative (yellow) magnetic polarities of sunspots. Equator is represented by the thick arc curve marked EW. These are the same regions as those in Fig. (\ref{tilt_it})\label{tilt_fin}}
\end{center} 
\end{figure}

The iteration continues with a new set of parameters: a non-zero tilt angle, a centroid of the total magnetic flux,  and centroids of positive and negative polarities; the long semi-major axis was calculated by newly determined parameters, 

\begin{equation}
a=3(r_p+r_n+s)
\end{equation}

\noindent where $r_p$ and $r_n$ are the radii of the circular areas of idealized positive and negative polarities; and $s$ is the separation of the bipole. The semi-minor axis, $b$, is the bigger of values between $r_p$ and $r_n$.  Successive tilt angles were compared, and the iteration is set to stop when the difference is less than or equal to $0^\circ.1$. Convergence takes $\le$20 iterations for $70$\% of sunspots while $\sim$80\% of spots converge to $0^\circ.1$ within 24 iterations.  Visual inspection confirmed that the tilt angle determinations are reasonable for even the largest number of iterations. The process is illustrated in Fig. (\ref{tilt_it}) where multiple ellipses overlapping a sunspot group on a magnetogram show successive solutions from the algorithm. The initial ellipses are evident by their zero orientation with respect to the equator. The fitted ellipses quickly turn to the final orientations during the process (black ellipses). The final derived  tilt angles of the sunspots in Fig (\ref{tilt_it}) are shown in Fig. (\ref{tilt_fin}). 
%Iteration of 24 times does not mean a bad tilt angle is produced. This only means that more runs are needed in order to reduce the tilt angle uncertainty to less than $0.1^\circ$. 

The fact that we start with identified sunspots in step (a) minimizes confusion in the identification of active regions on the magnetograms. Without anchoring the method with pattern recognition that originally came from human observers, the algorithm would have required a much more sophisticated method. 

%\section{Assessment of the Tilt Angle}
In our analysis, only sunspots having central meridional angle $\leq45^\circ$ were used for the tilt angle study in order to minimize errors caused by projection. This is equivalent to a 6-day period of a distinct sunspot group disk passage. The tilt angles were calculated with the automated IDL program for about 28642 sunspot groups observed by MWO from 1974 to 2012 March  and 11932 by MDI from 1996 to 2011 April. Although the MDI data have a shorter observing time span (1996-2011) compared to MWO data, MDI has higher observing rates than MWO because of the unlimited weather conditions and non-local time zone. 

About 20\% of sunspots in MWO data, and 12\% of sunspots in MDI data were considered to have poorly determined tilt angles, 
%(not convergent to $0^\circ.1$ within 24 iterations), 
and were visually re-examined after the automated program run. These sunspots were singled out by the following criteria: (a)  During the disk passage, those sunspot groups had standard deviations of tilt angles $\geq20^\circ.5$. This is equivalent to a tilt angle varying from $60^\circ$ to $120^\circ$ for a sunspot group in 6 days. We note that many sunspot groups have tilt angle variations greater than $20^\circ.5$ in the course of maximum 6 days, but yet the tilt angles appear valid. Therefore, $20^\circ.5$ is a low threshold to detect the ill-defined tilt angles; (b) Those sunspots having latitudes less than $15^\circ$ but the tilt angles greater than $ 45^\circ$. This ensures that a large tilt angle is valid for sunspots at low latitudes; (c) Those sunspots having bipolar separation greater than $15^\circ$. Based on the SWFC sunspot reports from 1991-2012, about only 1.5\% of sunspot groups have longitudinal extent greater than $15^\circ$. 
%Therefore, it is rare to have the distance between opposite polarities greater than $15^\circ$ in a sunspot group. 

For those sunspot groups with ill-defined tilt angles, we re-determined the tilt angles by manually entering the sunspot central positions, the semi-major and minor axes and the initial tilt angles. During the manual process, some sunspots were rejected altogether due to the poor observations of sunspots on a magnetogram or they were too close to other major sunspots. About 2000 sunspots were rejected from MWO data, and about $\sim150$ sunspots were rejected from MDI data.  These correspond to 7.5\% of MWO  and 1.2\% of MDI data samples, respectively.  Although we rejected these sunspots, the following results would not be materially altered by their inclusion.
%The final total numbers of sunspots having their tilt angles determined are 26716 for the MWO data, and 11780 for the MDI data. 

%Errors of the tilt angles are assumed a normal distribution. Based on the {\it Power Analysis}, the sample size of $3\times \sqrt{N}$  will ensure a statistical significant level 95\%,  where $N$ is the total number of sunspots. Randomly selected sunspots of 500 with MWO data, and 350 with the MDI data are visually examined for their magnetic tilt angle reasonability. 

\begin{figure}[t]
%\epsscale{1.2}
\begin{center}
%\plotone{tilt_def.png}
%\includegraphics[width=1.0\textwidth]{tilt2fit_mdi.png}
\includegraphics[width=1.0\textwidth]{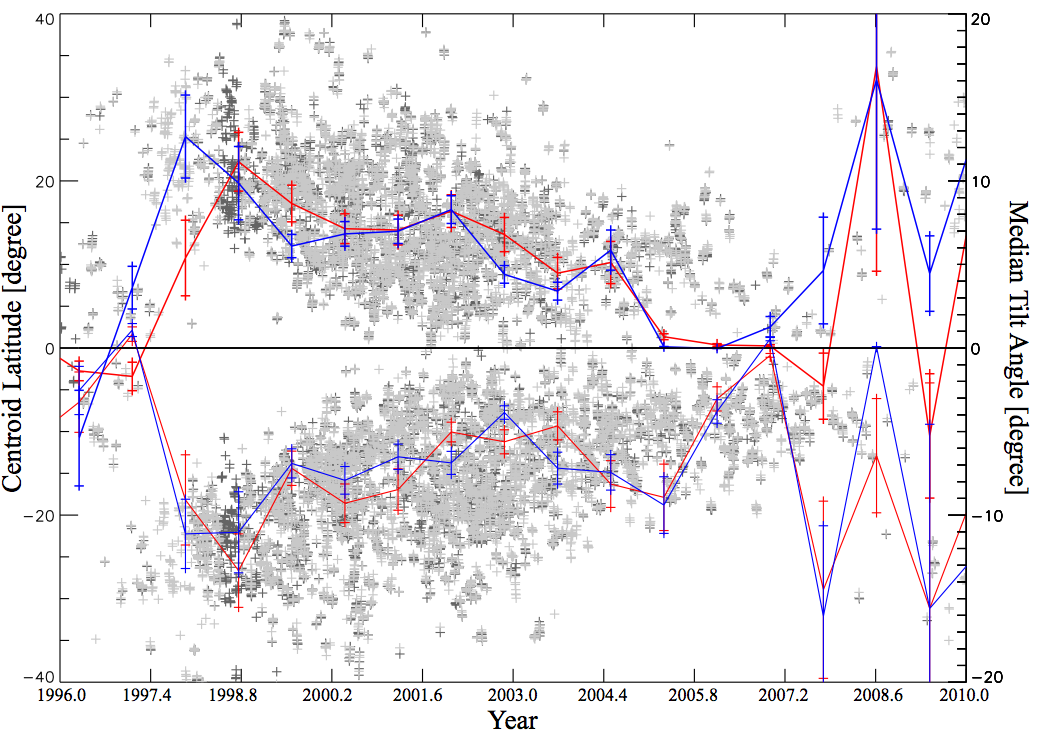}
\caption{Sunspot magnetic tilt angles were measured from MWO (red curve) and MDI (blue curve) magnetograms. They are over-plotted with latitudes of sunspot mass centers. The dark colored ``+'' represent sunspots measured with MWO, and the light colored ``+'' represent those with MDI.  The horizontal line represents the equator and $0^\circ$ tilt angle. The scales for the sunspot latitudes is to the left vertical axis, and scales for the tilt angles is to the right vertical axis.  The horizontal axis is time measured in ``year''.  \label{tilt2fit_mdi}}
\end{center} 
\end{figure}

The MWO and MDI data overlapped from 1996 to 2010, allowing us to compare results from these independent data sets.  We find that the sunspot magnetic tilt angle measurements agree remarkably well between two different instruments. This is shown in Fig. (\ref{tilt2fit_mdi}) where tilt angles from the two instruments are distinguished by colors, red (MW) and blue (MDI). The sunspot butterfly diagram is shown in the background in which latitudes are centroids of sunspot groups measured from MWO (dark colored ``+'') and MDI (light colored ``+''). Because of the agreements between tilt angle measurements from two instruments, we combine them for the time period from 1974 to 2012. Specifically, tilt angles presented in the following sections were measured from MWO data between 1974 to 1995, and after 2010. The tilt angles were measured from MDI between 1996 and 2010. MDI data gaps between June 1998 to January 1999 were filled with the tilt angle measurements from MWO. Over all, the total sunspot number used for the tilt angle study is 30623, and they made up 8705 distinct sunspot groups.

\section{Results}

\subsection{Hale's Law}
\begin{figure}[t]
%\epsscale{1.2}
%\begin{center}
%\plotone{tilt_def.png}
%\includegraphics[width=1.0\textwidth]{tilt2hale.png}
\includegraphics[width=1.0\textwidth]{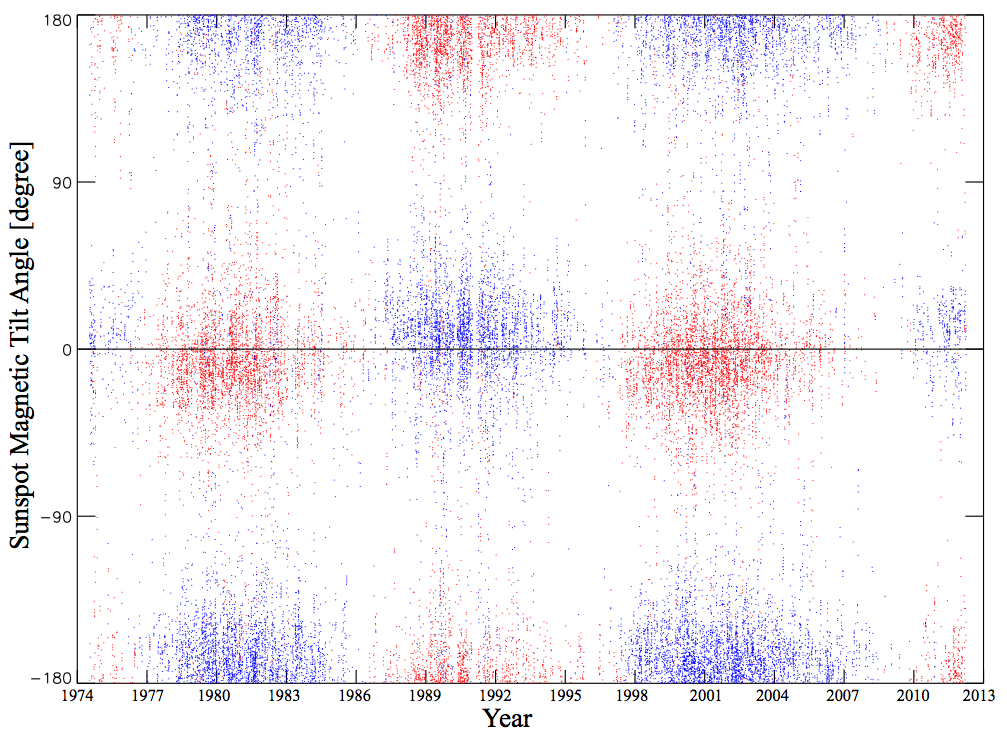}
\caption{Hale's law during 1974 to 2012. The time spans from the end of cycle 20 to the beginning of cycle 24, and is marked in year. The horizontal line represents the equator. Red dots represent sunspots in the northern hemisphere, and blue dots represent those in the southern hemisphere. \label{tilt_hale}}
%\end{center} 
\end{figure}

The tilt angles of all 30623 sunspots are presented in Fig. (\ref{tilt_hale}).  Hale's law is evident in that the majority of sunspots have opposite polarity distributions in the northern and southern hemispheres at any given cycle.  The leading polarity switches signs in the period when an old cycle fades, and a new cycle rises. We tabulate the percentages of sunspot numbers with their leading polarities either positive ($\gamma\leq 90^\circ$) or negative ($\gamma>90^\circ$) in the respective hemisphere from cycle to cycle. We adopt the official cycle definition available at the FTP site\footnote{ftp://ftp.ngdc.noaa.gov/STP/SOLAR\_DATA/SUNSPOT\_NUMBERS/INTERNATIONAL/maxmin/MAXMIN}. In our analysis, starts of cycles are those in the column ``Year of Minimum'', and the lengths of cycles are those listed in the column ``Cycle Length''. Three complete cycles were covered by our study, 21, 22, and 23.  Within each cycle, Table (\ref{hale}) shows that more than 90\% of sunspots had ``normal'' polarity alignment along the east-west directions. 

%%JING - THE LINK IS TOO LONG TO FIT AND SHOULD BE FIXED

\subsection{Joy's Law}
\begin{figure}[t]
%\epsscale{1.2}
%\begin{center}
%\plotone{tilt_def.png}
%\includegraphics[width=1.0\textwidth]{tilt_joy_his.png}
\includegraphics[width=1.0\textwidth]{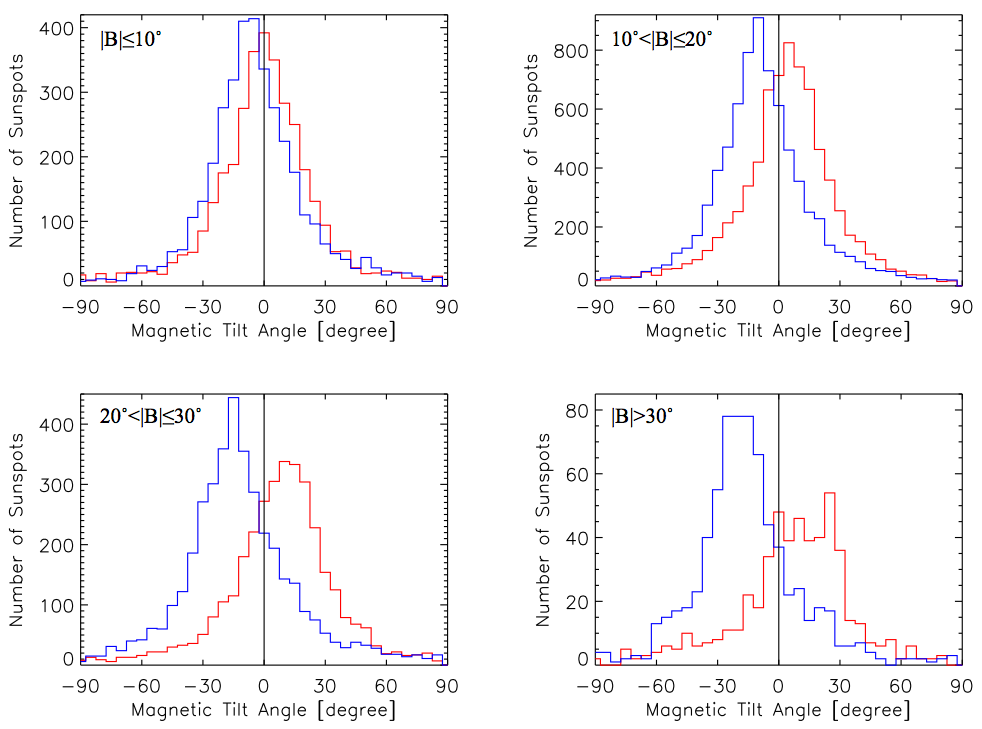}
\caption{Sunspot tilt angle distribution in northern (red lines) and southern (blue lines) hemispheres within four latitude ranges. The histogram bin size is $5^\circ$. The vertical straight lines mark the $0^\circ$ tilt angle (see Table \ref{tilt_summary}).\label{ssn2tilt}}
%\end{center} 
\end{figure}

Fig. (\ref{ssn2tilt}) shows histograms of the {\it sunspot tilt angle distributions} in four latitudinal zones in the northern (red) and southern (blue) hemispheres. The tilt angles were measured between [$-90^\circ, 90^\circ$] using the definition illustrated in Fig. (\ref{tilt_def}b). Within each latitude zone, the sunspot numbers peak at different tilt angles in the two hemispheres. The separation between the twin peaks increases with increasing latitude.

The median tilt angle, $\bar\gamma$, is the angle dividing the {\it sunspot tilt angle distribution function}, $n_s(\gamma)$, into equal areas. To compute $\bar\gamma$, we construct the {\it tilt angle cumulative distribution function} from $n_s(\gamma)$,

\begin{equation}
N_s(\gamma)=\int_{-90^\circ}^\gamma{n_s(\gamma)d\gamma}
\label{cumulative}
%\bar\gamma=F\left(\frac{1}{2}\max(\int{n_s(\gamma)d\gamma})\right)
\end{equation}

\noindent The median tilt angle is half the maximum cumulative function, 

\begin{equation}
\bar\gamma=F\left (\frac{1}{2}\max[N_s(\gamma)]\right )
\label{halfmax}
\end{equation}

\noindent where $F$ is the inverse function of $N_s(\gamma)$. The uncertainties of the median tilt angles were estimated from $3\times\bar\gamma/\sqrt{N_s}$, where $N_s$ is the total number of sunspots used to calculate the median tilt angle. Table (\ref{tilt_summary}) lists the median tilt angles calculated from the tilt angle distributions. It is consistent with what is shown in Fig.(\ref{ssn2tilt}). The median tilt angles clearly increase in absolute value with increasing latitude. 

A tilt angle hemispheric asymmetry is shown in Fig.(\ref{ssn2tilt}) and Table (\ref{tilt_summary}). The sunspot number peaks closer to $0^\circ$ tilt angle in the northern hemisphere than in the southern hemisphere. We found that the asymmetry in the tilt angles between the hemispheres existed in all cycles.  On average, the tilt angles were higher in the southern than in the northern hemisphere by $\sim5^\circ$ in the low latitude zone ($0^\circ-10^\circ$), $\sim10^\circ$ in the intermediate latitude zone ($10^\circ-30^\circ$)  and $\sim13^\circ$ in the high latitude zone ($>30^\circ$). 
%In addition, the observations also show tilt angles are widely variable at a given latitude. We use the quantity {\it r.m.s.} (Root-Mean-Square) to represent the variation of tilt angles, and the numbers are listed under {\it r.m.s.} in Table (\ref{tilt_summary}). We will discuss the {\it r.m.s.} of tilt angles in the Discussion section.
%{\it r.m.s.} are rather constant for different latitudinal zones and hemispheres.

\begin{figure}[t]
%\epsscale{1.2}
%\begin{center}
%\plotone{tilt_def.png}
%\includegraphics[width=1.0\textwidth]{tilt_joy.png}
\includegraphics[width=1.0\textwidth]{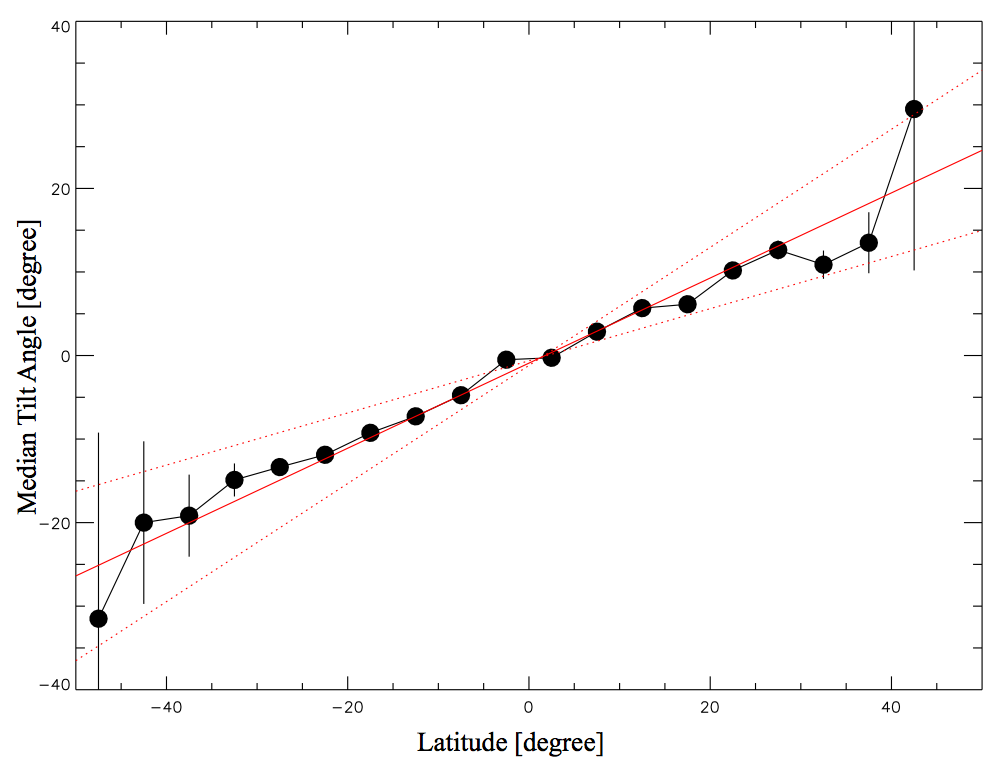}
\caption{Illustration of Joy's law. Circles and solid curve represent median tilt angles of all 30,600 sunspots calculated with Equations (\ref{cumulative}) and (\ref{halfmax}) in the $5^\circ$ latitude bins. The vertical bars represent uncertainties calculated with $3\times \gamma/\sqrt{ssn}$, where $ssn$ is the total sunspot number for the median $\gamma$. The solid and dotted red lines are the linear least square fits to the data points and their uncertainties respectively. Joy's law is expressed as $\bar\gamma=(0.5\pm0.2)B-(0.9\pm0.3)$, where $B$ is the latitude.   \label{tilt_joy}}
%\end{center} 
\end{figure}

Joy's law is also illustrated in Fig. (\ref{tilt_joy}) in which median tilt angles were calculated using Equations (\ref{cumulative}) and (\ref{halfmax}) within $5^\circ$-latitudinal bins. The linear correlation between the sunspot latitude, $B$, and the median tilt angle is written $\bar\gamma(B)=k B +c$. Both constants $k$ and $c$ are to be determined by fitting data points (black circles in Fig (\ref{tilt_joy})). The linear least-square fit produces the red straight line,

\begin{equation}
\bar\gamma(B)=(0.5\pm 0.2)B-(0^\circ.9\pm 0^\circ.3).
\label{joy}
\end{equation}

%JING: YOU SHOULD MENTION PRIOR WORK ABOUT THE 0.5 - DIDN"T SOMEONE CONNECT IT WITH CORIOLIS FORCE SOMEHOW?  JUST MNTION IT IN A SENTENCE>  YOU CAN STILL HAVE ANOTHER PAPER TO DIG IN MORE DETAIL ABOUT IT

\noindent The uncertainty of the slope was propagated from the median tilt angle uncertainties. Two red straight dotted lines are fits to the median tilt angles $\pm$ error bars. The differences between two fits gave the errors of $3\sigma$, which are $\pm 0.2$ for Joy's law slope, and $\pm0^\circ.3$ for the tilt angle at $B=0^\circ$. The tilt angle hemispheric asymmetry discussed above is the reason why the constant term, $c=-0^\circ.9$, but not zero. 

\citet{1991ApJ...375..761W} gave an empirical equation for tilt angles by assuming that the Coriolis force acts on an emerging flux rope as it undergoes expansion in the longitudinal direction while it is below the sun's surface. Their study was based on examining  2700 bipolar magnetic regions (BMRs) erupted between 1976-1986 \citep{1989SoPh..124...81W}. The tilt angle, $\gamma$, of the flux tube follows $\sin\gamma \sim \omega(B)\tau_{ex} \sin B$, where $\omega(B)$ is the differential rotation rate as a function of latitude $B$, $\tau_{ex}$ represents the sub-surface flux rope expansion time. For a typical solar rotation rate $14^\circ.4$/day, and $\tau_{ex}=2.0$ days, we obtain $\omega(B)\tau_{ex}=0.5$ which is the correlation coefficient between $\sin\gamma$ and $\sin B$. Coincidently, this agrees very well with our measurements: $\sin\bar\gamma=(0.5\pm0.2)\sin B$ when $\sin\bar\gamma$ and $\sin B$ are fitted with a linear least  square fit.

% It is not clear why such hemispheric asymmetry exists except for that the southern hemisphere had more sunspots than in the northern hemisphere by 7\% on average. The percentage increased with increasing latitude (see Table (\ref{tilt_summary})). 
%%JING: WHAT DOES THE LAST SETENCE MEAN?  THE SUBJECT IS AMBIGUOUS

%For Joy's law cycle by cycle, the fitted $k$ and $c$ have bigger uncertainties than the Joy's law from the combine measurements described above. They are listed in Table (\ref{joy_cycle}) for three complete cycles.  
%It seems that the Coriolis force can statistically explain the tilt angles.

\subsection{Tilt Angles  and Sunspot Butterfly Diagram}
\begin{figure}[t]
%\epsscale{1.2}
%\begin{center}
%\plotone{tilt_def.png}
%\includegraphics[width=1.0\textwidth]{tilt2fit.png}
\includegraphics[width=1.0\textwidth]{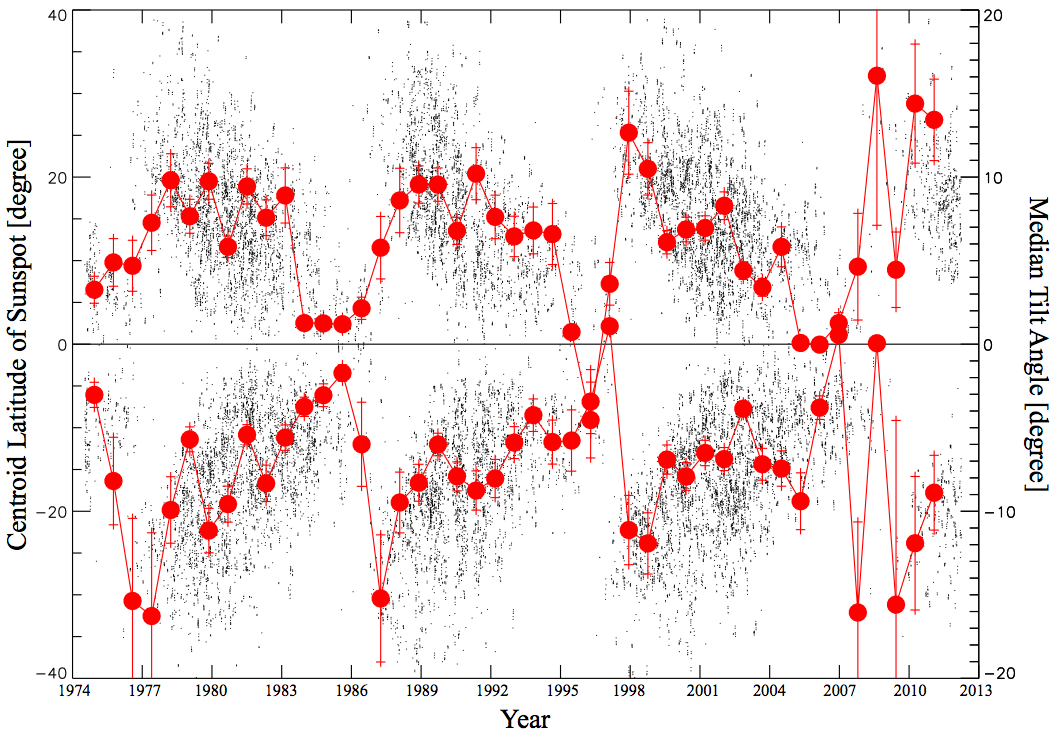}
\caption{Median tilt angles (red curves) varying with time [year] overlapping the butterfly diagram.  The median tilt angles were calculated with Equations (\ref{cumulative}) and (\ref{halfmax}) in the 300 day interval. The uncertainties were estimated as $3\times \bar\gamma/\sqrt{ssn}$, where $ssn$ is the number of sunspots which contributed to the median tilt angle. The butterfly diagram is made with centroids of sunspots. The latitudes are scaled in the left vertical axis, and the tilt angles are scaled in the right vertical axis. The horizontal straight line indicate the latitude and tilt angle $0^\circ$. \label{tilt_fit}}
%\end{center} 
\end{figure}

The average latitude of sunspots depends on the phase of the sunspot cycle as illustrated by the butterfly diagram. The tilt angles may also vary with time due to Joy's law. In order to determine if the time dependence of the tilt angle is strictly a consequence of the sunspot changing latitudes, we calculated median tilt angles, $\bar\gamma$, using the histogram-cumulative function technique described by Equations (\ref{cumulative}) and (\ref{halfmax}). The sunspot {\it tilt angle distribution function} is constructed in a series of time intervals, $t-\Delta t/2$ and $t+\Delta t/2$, where $\Delta t=150$ days in our analysis. The measurement uncertainties were calculated, again, as $3\sigma$, where $\sigma=\bar\gamma/\sqrt{N_s}$. Fig. (\ref{tilt_fit}) shows the median tilt angles as a function of time overlapping a butterfly diagram made from the centroids of the sunspot latitudes determined from the total magnetic flux. The latitude scale for sunspots is on the left vertical axis, and  the tilt angle scale is on the right vertical axis. The median tilt angles are plotted as red curves, and circles. It appears that tilt angles decrease with time in both hemispheres as the sunspots migrate from high to low latitudes through each activity cycle.

\begin{figure}[t]
%\epsscale{1.2}
%\begin{center}
%\plotone{tilt_def.png}
%\includegraphics[width=1.0\textwidth]{tilt_ssn.png}
\includegraphics[width=1.0\textwidth]{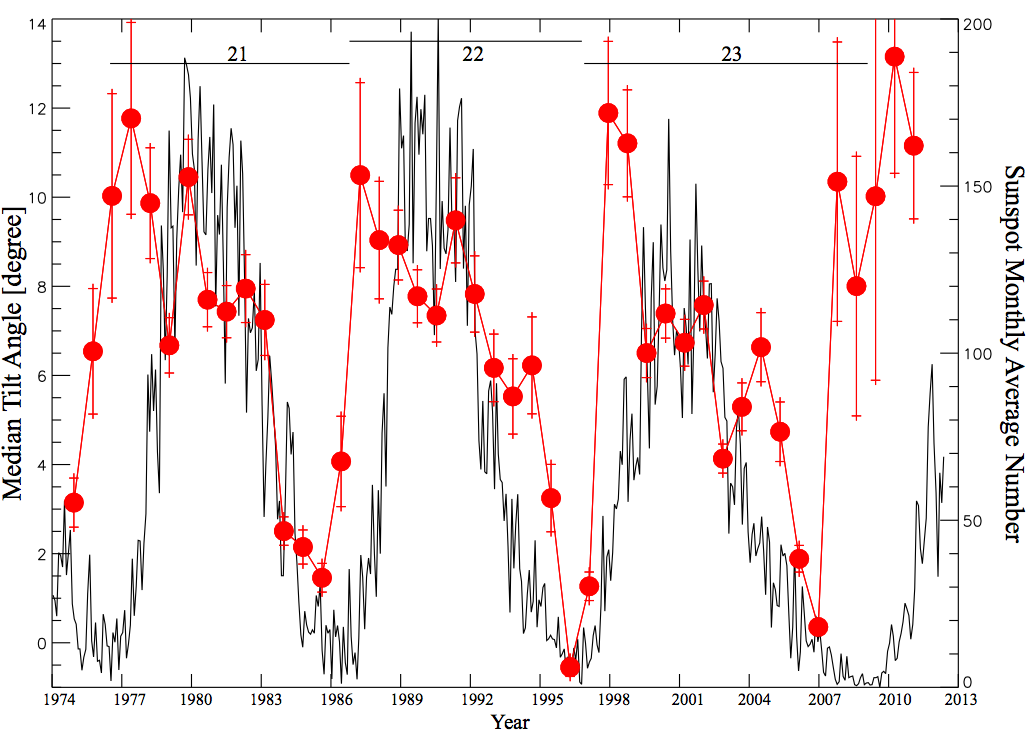}
\caption{Median magnetic tilt angles (red curve and circles) and sunspot monthly average numbers (black curve) as functions of time [year]. The tilt angles were combined from both hemispheres (shown in Fig. \ref{tilt_fit}), and uncertainties were calculated accordingly. The scale for tilt angles is to the left vertical axis, and the scale for the sunspot numbers is to the right vertical axis. Three horizontal bars on top of the plot represent the starts and durations of cycles 21, 22, and 23. The starting dates and durations are listed in Table (\ref{hale}). \label{tilt_ssn}}
%\end{center} 
\end{figure}

\begin{figure}[t]
%\epsscale{1.2}
%\begin{center}
%\plotone{tilt_def.png}
%\includegraphics[width=1.0\textwidth]{tilt2cycle.png}
\includegraphics[width=1.0\textwidth]{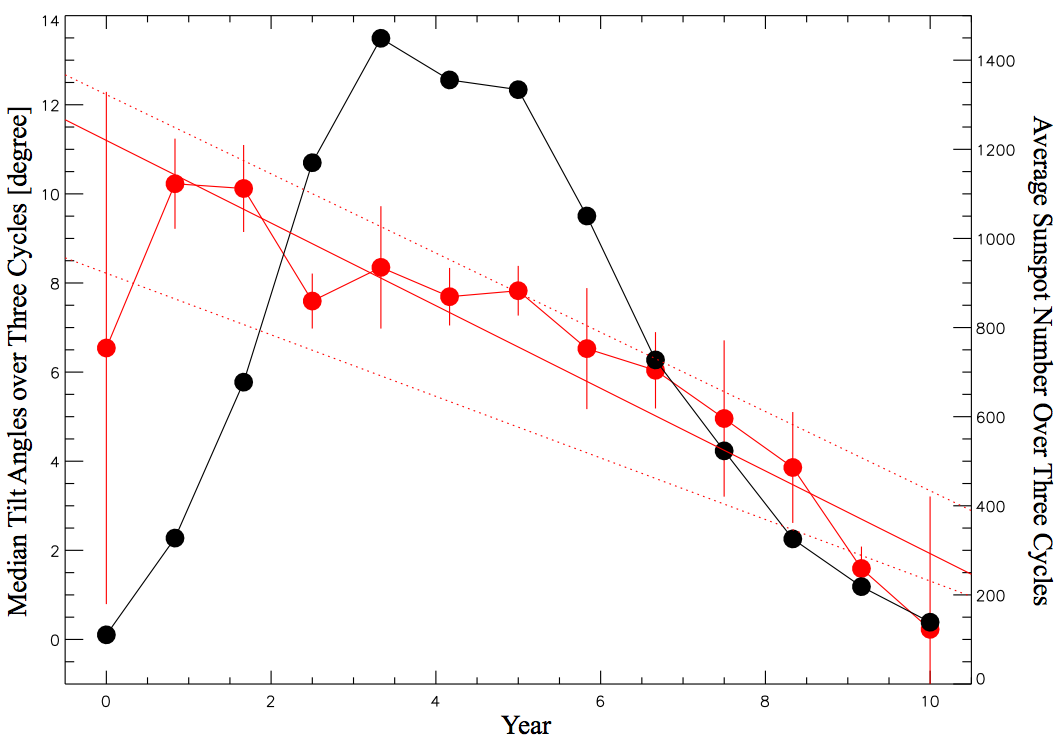}
\caption{The median tilt angles (red) and average sunspot numbers (black) as functions of time (year) within a typical cycle. The tilts and sunspot numbers are averaged over three cycles, 21, 22, and 23 which were registered at the respective cycle minima with equal time intervals 3652 days ( $\sim10$ yrs). The uncertainties of the tilt angles were the standard deviation of tilt angles at each data points.  The tilt angles are fitted with the error-weight linear least square fit (solid red line) $\gamma=(-0^\circ.9\pm0^\circ.1)t+(11^\circ.2\pm2^\circ.0)$, where $t$ is measured in year from the beginning of the solar cycle. Two red dotted lines represent the uncertainties of fittings.  \label{tilt_cycle}}
%\end{center} 
\end{figure}

Fig (\ref{tilt_ssn}) shows median tilt angles over the monthly averaged sunspot number (available at http://solarscience.msfc.nasa.gov/greenwch/spot\_num.txt). The tilt angles were averaged between two hemispheres shown in Fig (\ref{tilt_fit}) in order to increase the signal-to-noise. The uncertainties were calculated accordingly. The fact that tilt angles show the general declining trend with cycle phases allows us to combine the tilt angle measurements of cycles 21, 22 and 23. We registered tilt angle measurements at times of cycle minima with equal time intervals, 3652 days ($\sim 10 $ years). This is illustrated by three horizontal bars on top of figure (\ref{tilt_ssn}). With improved signal-to-noise levels, Fig (\ref{tilt_cycle}) shows the combined tilt angles as a function of time within a typical cycle. Error bars represent $\pm1$ standard deviations of tilt angles of three cycles. Large error bars of tilt angles from year 0 to 1 were caused by fewer sunspots and their diverse tilt angles during the cycle exchange period.  By fitting tilt angles (red circles) with an error-weighted linear least-square fit, we obtain

%We register the highest points of tilt angles from three cycles and combine them in the time range 250 days before and 3600 days after the highest tilt angle points. Three equal time intervals, 3850 days ($\sim10.54$ years) in total, are indicated by horizontal bars above the sunspot number curves in Fig. (\ref{tilt_ssn}). 
\begin{equation}
\bar\gamma=(-0^\circ.9\pm 0^\circ.1) ~t+(11^\circ.2\pm2^\circ.0)
\label{cycle}
\end{equation}

\noindent where $t$ (year) is the time since the start of the solar cycle. The errors of averaged tilt angles produced the two red dotted lines. The differences between the lines are the uncertainties in Equation (\ref{cycle}). On average, the tilt angles decrease $\sim0^\circ.9$ per year. Within a typical 11-year solar cycle, the tilt angles decrease about $10^\circ$. 

\begin{figure}[t]
%\epsscale{1.2}
%\begin{center}
%\plotone{tilt_def.png}
%\includegraphics[width=1.0\textwidth]{latitude2cycle.png}
\includegraphics[width=1.0\textwidth]{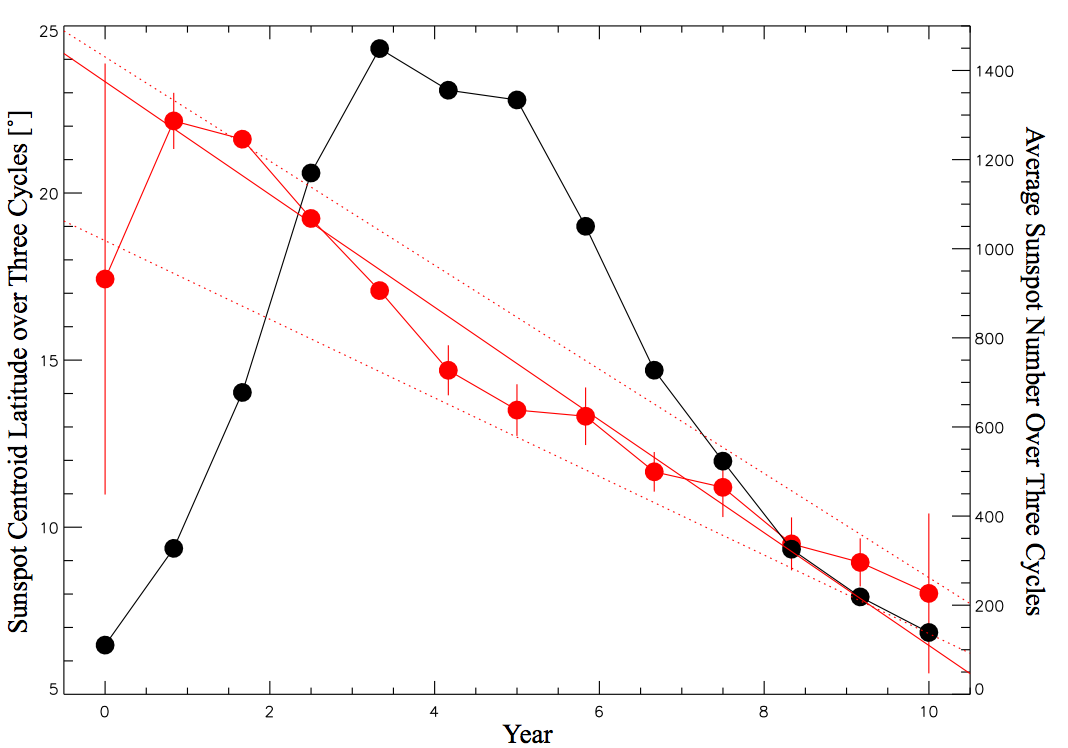}
\caption{The sunspot centroid latitude (red) and average sunspot numbers (black) as functions of time (year) within a typical cycle. The tilts and sunspot numbers are averaged over three cycles, 21, 22, and 23 which is the processed in the same way as those in Fig. (\ref{tilt_cycle}). The uncertainties were 1.5 times standard deviations of latitudes at each data points.  The linear least square error-weighted fit is plotted by the solid red line, and two dotted lines represent the uncertainties of fittings. The relation is given by  $B=(-1^\circ.7\pm0^\circ.2)t+(23^\circ.3\pm2^\circ.7)$, where $t$ is measured in year from the beginning of the solar cycle.   \label{lat2cycle}}
%\end{center} 
\end{figure}

The latitudes of sunspot centroids averaged from three complete cycles are shown in Fig (\ref{lat2cycle}). It is consistent with the butterfly diagram that the sunspots generally migrate from high to low latitudes as the cycle progresses. The median sunspot latitudes and their uncertainties as a function of  time were linearly fitted: $\bar B(t)=(-1^\circ.7\pm 0^\circ.2)t+(23^\circ.3\pm 2^\circ.7$), where $t$ is measured in year from the beginning of the cycle. On average, sunspot groups migrate from high to low latitudes at the rate $\sim 1^\circ.7$ yr$^{-1}$, while the tilt angle decreases $0^\circ.5$ per degree of latitude (see Equation (\ref{joy})). This results in the tilt angle decreasing roughly $0^\circ.9$ yr$^{-1}$ which rate is consistent with Equation (\ref{cycle}). This implies that sunspot tilt angles decrease with time following Joy's law.

\subsection{Tilt Angles and Cycle Phases}
\begin{figure}[t]
%\epsscale{1.2}
%\begin{center}
%\plotone{tilt_def.png}
%\includegraphics[width=1.0\textwidth]{tilt2cycle_3B.png}
\includegraphics[width=1.0\textwidth]{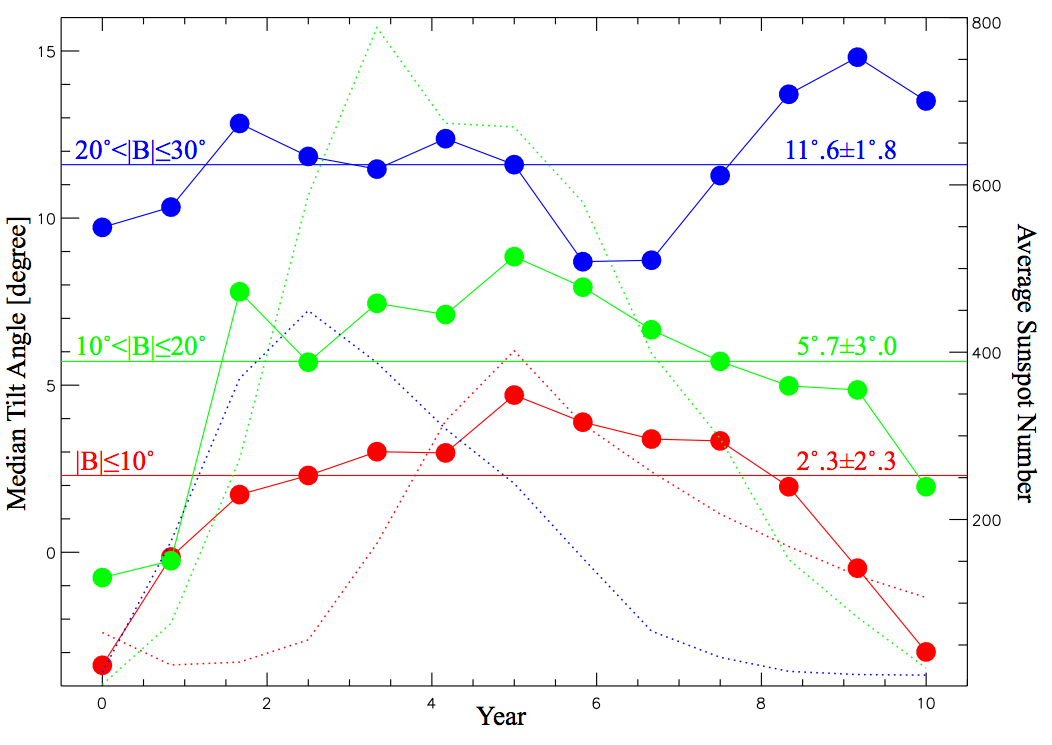}
\caption{Median tilt angles (solid curves, and circles) and average sunspot numbers (dotted curves) within three latitude zones as functions of time. The scale for the tilt angles is to the left vertical axis, and the scale for the sunspot numbers is to the right vertical axis. The tilts are averaged over three cycles, 21, 22, and 23. Three average tilt angles (colored horizontal lines) and their standard deviation at the respective latitude ranges are written above the colored lines.  Colors represents latitude zones. Red:  $|B|\leq 10^\circ$; Green: $10^\circ<|B|\leq 20^\circ$; and Blue: $20^\circ<|B|\leq 30^\circ$.  \label{tilt3B}}
%\end{center} 
\end{figure}

Within a limited latitude range, the sunspot tilt angles were approximately invariant with respect to time. Fig. (\ref{tilt3B}) shows the tilt angles averaged over three cycles within three latitude zones as functions of time. Spots with $|B|>30^\circ$ are not plotted because there are too few of them, and the uncertainties are too great. The data points are represented by colored circles and solid curves. Joy's law is, again, evident in that the median tilt angles are generally higher as the latitude zones increase. The blue curve connecting blue circles represent the sunspots at latitudes $20^\circ<|B|\leq 30^\circ$ and stays generally at the highest tilt angle range.  The green curve connecting green circles represent sunspots at latitudes $10<|B|\leq 20^\circ$ and occupies the intermediate tilt angle range. The red curve connecting red circles represent sunspots at latitudes $|B|\leq 10^\circ$ and outlines the low tilt angles.  Within the tilt angle uncertainties (not shown in the figure), the figure shows that tilt angles stayed roughly constant through time at a fixed latitude zone. We plot the median tilt angles with colored horizontal lines corresponding to each latitudinal zone. Above respective lines, the median tilt angles and their uncertainties are marked with respective colors. The sunspot tilt angles are independent of cycle phases. 
%This seems to contradict the Fig. (\ref{tilt_cycle}) and the Equation (\ref{cycle}) which show the median tilt angles decrease with time.  

The seeming contradiction between observations in Fig. (\ref{tilt_cycle}) or Equation (\ref{cycle}) and Fig. (\ref{tilt3B}) can be understood by numbers of sunspots varying with cycle phases and latitudes. In Fig. (\ref{tilt3B}), three color-dotted curves represent sunspot numbers used to calculate the tilt angles at three different latitudinal zones (the scale is shown by the right vertical axis). On average, there were fewer sunspots at low latitude (red and green dotted curves) than at high latitude (blue dotted blue curve) in the beginning of the cycle (around year=1). This brings the median tilt angles calculated from Equations (\ref{cumulative}) and (\ref{halfmax}) to higher values. Likewise, the red and green dotted curves rise and  the blue dotted curve declines with time. As the result, the median tilt angles decrease as sunspots migrate from high to low latitude as cycle progresses, but the median tilt angles are independent of cycle phases at a fix latitudinal range.

\section{Summary}
We calculated the sunspot magnetic tilt angles using daily average magnetograms by the Mt. Wilson Observatory from 1974 to 2012, and MDI magnetorams from 1996 to 2010.  The tilt angles were measured from 30650 sunspots. We summarize the statistical results of the sunspot tilt angles in the time period 38 years:

\begin{enumerate}

%\item The total magnetic flux, $\Phi$, varying with time agree with the cycle variations. The total unsigned magnetic flux are correlated with the magnetic area, $\Phi=2.0\times10^{-7}a_m^{1.445}$ Maxwells, where $a_m$ is the sunspot magnetic area measured in [cm$^2$]. The total magnetic flux has more complicated relation with the pole separation than a simple linear correlation.

\item Hale's law is evident in the measured sunspot magnetic tilt angles. Within each cycle, over 90\% of sunspot polarity follow  Hale's law in both hemispheres, and sign alignment is reversed by the start of each new cycle.

\item On average,  Joy's law is evident in that the tilt angles increase with increasing latitude following the relation $\bar\gamma(B)=0.5 B-0^\circ .9$, where $B^\circ$ is the latitude. 

%\item The Coriolis force is responsible for the median tilt angles following Joy's law. However, the Coriolis force does not produce the wide variance of the tilt angles, which has a general $r.m.s. \sim 27^\circ$. 

\item On average, the tilt angles decrease with time at the rate $0^\circ.9$ per year. This largely reflects Joy's law following the sunspot butterfly diagram. 

\item Within latitudinal zones, the tilt angles were independent of the cycle phases.  The average tilt angles are $2^\circ$ in the low latitudinal zone ($|B|\leq 10^\circ$), $6^\circ$ in the median-low latitudinal zone ($10^\circ<|B|\leq 20^\circ$) and $12^\circ$ in the median-high latitudinal zone ($|B|>30^\circ$).

%\item The trend of magnetic tilt angles is independent of the cycle strength measured either in magnetic flux or in sunspot numbers. 

\item We found an unexplained, persistent  asymmetry between the median tilt angles measured in the north and south hemispheres in all latitudinal ranges. The average tilt angles were greater in the southern hemisphere than  in the northern hemisphere by $\sim$6$^\circ$ in the 0 to $10^\circ$ latitude zone, by $\sim$10$^\circ$ in the intermediate latitudel zone ($10^\circ-30^\circ$) and by $\sim$13$^\circ$ in the high latitude zone ($>30^\circ$). 

%\item The unusual behavior of the recent solar minimum was reflected in the tilt angle measurements. The tilt angles experienced dramatic variation during the prolonged minimum period of the cycle 23.

\end{enumerate}

\acknowledgments
We thank D. Jewitt, P. Gilman and the anonymous referee for comments which greatly improved the quality of the paper. The synoptic program at MWO has benefited from support by NASA, NSF and ONR over many years.  The long record of magnetic field data would not have been possible without support from these agencies.  Continuing observations depend on current support which comes from NSF through grant AGS-0958779, and NASA through grants NNX09AB12G and HMI subcontract 16165880.  
%$\omega=a+b\sin^2\bar B+c\sin^4 \bar B$ where $a=13^\circ.76$ day$^{-1}$; $b=-1^\circ.74$ day$^{-1}$; and $c=-2^\circ.19$ day$^{-1}$
%
%

\clearpage

\begin{deluxetable}{ccccccc}
\tablenum{1}
%\tabletypesize{\scriptsize}
\tablecolumns{7}
\tablecaption{Hale's Law: Leading spots with positive magnetic polarities \label{hale}}
\tablewidth{0pt}
\tablehead{
\colhead{Cycle (Length)} &
\colhead{Year of} &
\multicolumn{2}{r}{Northern Hemisphere} &
\colhead{} &
\multicolumn{2}{r}{Southern Hemisphere} \\
\cline{3-4} \cline{6-7} \\ 
\colhead{[yrs]}&
\colhead{Minimum} &
\colhead{SSN} &
\colhead{$\gamma\leq 90^\circ$} &
%\colhead{$\gamma>90^\circ$} &
\colhead{} &
\colhead{SSN}&
\colhead{$\gamma\leq 90^\circ$} 
%\colhead{$\gamma>90^\circ$} 
}
\startdata
21 (10.3) & 1976.5 & 4425 & 91.5\% & & 4655 & 8.7\% \\
22 (10.0) & 1986.8 & 3781 & 9.1\%    && 4130 & 91.5\% \\
23 (12.2) & 1996.9 & 5457 & 92.3\% & & 6272  & 6.5\% \\
%23 & (MDI)&  5135 & 92.7\% & 7.3\% & 5783 & 6.6\% & 93.4\%\\
\enddata
%\tablenotetext{a}{Cycles are defined at \footnote{ftp://ftp.ngdc.noaa.gov/STP/SOLAR\_DATA/SUNSPOT\_NUMBERS/INTERNATIONAL/maxmin/MAXMIN}}
%%JING - THIS LINK IS TOO LONG TO FIT
\tablecomments{``SSN'' represents the total sunspot number used in the calculation of tilt angles in the respective categories. $\gamma$ is the tilt angle. Based on the tilt angle definition illustrated in Fig. (\ref{tilt_def}a), the leading spots have positive magnetic polarities when $\gamma\leq 90^\circ$.}
\end{deluxetable}

\clearpage

\begin{deluxetable}{clrlr}
\tablenum{2}
%\tabletypesize{\scriptsize}
\tablecolumns{5}
\tablecaption{Latitude Dependence of Tilt Angles (1974-2012) \label{tilt_summary}}
\tablewidth{0pt}
\tablehead{
\colhead{Latitude} &
\multicolumn{2}{c}{Northern Hemisphere} &
\multicolumn{2}{c}{Southern Hemisphere} \\
\colhead{Zone} &
\colhead{SSN} &
\colhead{$\bar\gamma\pm 3\sigma$} &
\colhead{SSN} &
\colhead{$\bar\gamma\pm 3\sigma$} 
}
\startdata
$\leq10^\circ$ &3449 & $-2^\circ.2\pm 0^\circ.1$  &3669 & $-8^\circ.3 \pm 0^\circ.4 $\\
$10^\circ - 20^\circ$ & 7442&$2^\circ.1\pm 0^\circ.1$ &7572& $-12^\circ.2\pm 0^\circ.4$ \\
$20^\circ - 30^\circ$ & 3374& $7^\circ.0 \pm 0^\circ.4$  &3888&  $-16^\circ.5\pm 0^\circ.8$\\
$>30^\circ$ & 519 & $7^\circ.0 \pm 0^\circ.9$ & 710 & $-19^\circ.9 \pm 2^\circ.2$\\
\enddata
%\tablecomments{Uncertainty of the median tilt angle, $\bar\gamma$, was estimated by the signal-to-noise principle, $\sigma=\pm \bar\gamma/\sqrt{ssn}$, where $ssn$ is the total sunspot number.}
%\tablecomments{SSN has the same meaning as those in Table (\ref{hale}).}
\end{deluxetable}

%\begin{figure}[t]
%%\epsscale{1.2}
%%\begin{center}
%%\plotone{tilt_def.png}
%\includegraphics[width=1.0\textwidth]{magflux2time.png}
%\caption{Total magnetic flux as a function of time. The times are in years.  The colored brightness represents the average sunspot numbers per day per Maxwells. The sunspot number density was calculated in the time interval 160 days. The brightness scale is shown in the lower right of the plot representing the sunspot numbers: 6.0, 4.0, 2.5, 1.0, 0.5, 0.2 from left to right. The numbers above each bright blobs are the numbers of the solar cycles. The magnetic fluxes were measured from MWO data.\label{flux2time}}
%%\end{center} 
%\end{figure}

%\begin{figure}[t]
%%\epsscale{1.2}
%%\begin{center}
%%\plotone{tilt_def.png}
%\includegraphics[width=1.0\textwidth]{magflux2area.png}
%\caption{Total magnetic flux as a function of the sunspot magnetic area. All sunspots observed by MWO are plotted in light colored ``+''. The contour represent sunspot number density per cm$^2$ per Maxwells. The contours are the sunspot numbers ($10^3$): 2.4, 4.8, 9.6, 15, 24, 30, and 36. The solid straight line is the linear fit between the logarithm magnetic flux, $\Phi$, and the logarithm magnetic area, $a_m$, of all the sunspots, $\log_{10}\Phi=1.4866\log_{10} a_m-7.5799$. \label{flux2area}}
%%\end{center} 
%\end{figure}

%\begin{figure}[t]
%%\epsscale{1.2}
%\begin{center}
%%\plotone{tilt_def.png}
%\includegraphics[width=1.0\textwidth]{tilt_def2.png}
%\caption{Illustration of the tilt angles vary in the range [$-90^\circ,90^\circ$]. Orientations of the bipole tilt angles are identical to those defined in Fig. \ref{tilt_def1}, but the polarity distributions of bipoles are no longer revealed by tilt angles. The tilt angle $\gamma \geq0$ when the leading pole lies equatorward of trailing pole, $\gamma<0$ when the trailing pole equatorward of the leading pole \citep{1991ApJ...375..761W}. \label{tilt_def2}}
%\end{center} 
%\end{figure}

%\begin{figure}[t]
%%\epsscale{1.2}
%%\begin{center}
%%\plotone{tilt_def.png}
%\includegraphics[width=1.0\textwidth]{magflux2s.png}
%\caption{Total magnetic flux as a function of pole separation. All sunspots are plotted in the symbols ``+''. The lighter colored ``+'' represent sunspot population having $d_e>0.435$ (see Equation (\ref{de})). The contours are the sunspots numbers to highlight the population ($10^3$): 2.4, 4.8, 9.6, 15, 24, 30, 36, and 42. The solid straight line is the linear least square fit to this population, $\log_{10}\Phi=2.34\log_{10} s+20.41$. The dotted line is a fit to the entire sunspot population,  $\log_{10}\Phi=0.83\log_{10} s+21.81$.\label{flux2s}}
%%\end{center} 
%\end{figure}

%\begin{figure}[t]
%%\epsscale{1.2}
%%\begin{center}
%%\plotone{tilt_def.png}
%\includegraphics[width=1.0\textwidth]{tilt2rms_histogram.png}
%\caption{Histogram of sunspot group numbers versus the {\it r.m.s.} of tilt angles of sunspot groups. {\it r.m.s.} of tilt angles of 2060 sunspot groups were calculated.  \label{tilt2rms}}
%%\end{center} 
%\end{figure}


\clearpage
\begin{thebibliography}{}
\bibitem[Babcock(1953)]{1953ApJ...118..387B} Babcock, H.~W.\ 1953, \apj, 118, 387
\bibitem[Babcock(1961)]{1961ApJ...133..572B} Babcock, H.~W.\ 1961, \apj, 133, 572 
\bibitem[Basu \& Antia(2010)]{2010ApJ...717..488B} Basu, S., \& Antia, H.~M.\ 2010, \apj, 717, 488
\bibitem[Cameron et al.(2010)]{2010ApJ...719..264C} Cameron, R.~H., Jiang, J., Schmitt, D., \& Sch{\"u}ssler, M.\ 2010, \apj, 719, 264  
\bibitem[Choudhuri et al.(1995)]{1995A&A...303L..29C} Choudhuri, A.~R., Schussler, M., \& Dikpati, M.\ 1995, \aap, 303, L29 
\bibitem[Dasi-Espuig et al.(2010)]{2010A&A...518A...7D} Dasi-Espuig, M., Solanki, S.~K., Krivova, N.~A., Cameron, R., \& Pe{\~n}uela, T.\ 2010, \aap, 518, A7  
\bibitem[D'Silva \& Choudhuri(1993)]{1993A&A...272..621D} D'Silva, S., \& Choudhuri, A.~R.\ 1993, \aap, 272, 621
\bibitem[Fan et al.(1994)]{1994ApJ...436..907F} Fan, Y., Fisher, G.~H., \& McClymont, A.~N.\ 1994, \apj, 436, 907 
\bibitem[Fisher et al.(1995)]{1995ApJ...438..463F} Fisher, G.~H., Fan, Y., \& Howard, R.~F.\ 1995, \apj, 438, 463  
\bibitem[Freeland \& Handy(1998)]{1998SoPh..182..497F} Freeland, S.~L., \& Handy, B.~N.\ 1998, \solphys, 182, 497 
\bibitem[Hale et al.(1919)]{1919ApJ....49..153H} Hale, G.~E., Ellerman, F., Nicholson, S.~B., \& Joy, A.~H.\ 1919, \apj, 49, 153 
\bibitem[Hale \& Nicholson(1925)]{1925ApJ....62..270H} Hale, G.~E., \& Nicholson, S.~B.\ 1925, \apj, 62, 270 
\bibitem[Hathaway \& Rightmire(2010)]{2010Sci...327.1350H} Hathaway, D.~H., \& Rightmire, L.\ 2010, Science, 327, 1350 
\bibitem[Howard et al.(1983)]{1983SoPh...87..195H} Howard, R., Boyden, J.~E., Bruning, D.~H., et al.\ 1983, \solphys, 87, 195 
\bibitem[Howard et al.(1983)]{1983SoPh...83..321H} Howard, R., Adkins, J.~M., Boyden, J.~E., et al.\ 1983, \solphys, 83, 321 
\bibitem[Howard et al.(1984)]{1984ApJ...283..373H} Howard, R., Gilman, P.~I., \& Gilman, P.~A.\ 1984, \apj, 283, 373 
\bibitem[Howard(1991a)]{1991SoPh..132...49H} Howard, R.~F.\ 1991, \solphys, 132, 49 
\bibitem[Howard(1991b)]{1991SoPh..132..257H} Howard, R.~F.\ 1991, \solphys, 132, 257
\bibitem[Howard(1991c)]{1991SoPh..136..251H} Howard, R.~F.\ 1991, \solphys, 136, 251
\bibitem[Howard(1994)]{1994SoPh..149...23H} Howard, R.~F.\ 1994, \solphys, 149, 23 
\bibitem[Gilman \& Howard(1986)]{1986ApJ...303..480G} Gilman, P.~A., \& Howard, R.\ 1986, \apj, 303, 480  
\bibitem[Guo et al.(2007)]{2007A&A...462.1121G} Guo, J., Zhang, H.~Q., \& Chumak, O.~V.\ 2007, \aap, 462, 1121
\bibitem[Komm et al.(2011)]{2011JPhCS.271a2077K} Komm, R., Howe, R., Hill, F., Gonz{\'a}lez Hern{\'a}ndez, I., \& Haber, D.\ 2011, Journal of Physics Conference Series, 271, 012077  
\bibitem[Kosovichev \& Stenflo(2008)]{2008ApJ...688L.115K} Kosovichev, A.~G., \& Stenflo, J.~O.\ 2008, \apjl, 688, L115 
\bibitem[Labonte \& Howard(1982)]{1982SoPh...80..373L} Labonte, B.~J., \& Howard, R.\ 1982, \solphys, 80, 373 
\bibitem[Leighton(1969)]{1969ApJ...156....1L} Leighton, R.~B.\ 1969, \apj, 156, 1 
\bibitem[Muneer \& Singh(2002)]{2002SoPh..209..321M} Muneer, S., \& Singh, J.\ 2002, \solphys, 209, 321 
\bibitem[Newton \& Nunn(1951)]{1951MNRAS.111..413N} Newton, H.~W., \& Nunn, M.~L.\ 1951, \mnras, 111, 413
\bibitem[Parker(1955a)]{1955ApJ...121..491P} Parker, E.~N.\ 1955, \apj, 121, 491 
\bibitem[Parker(1955b)]{1955ApJ...122..293P} Parker, E.~N.\ 1955, \apj, 122, 293 
\bibitem[Sattarov et al.(2002)]{2002ApJ...564.1042S} Sattarov, I., Pevtsov, A.~A., Hojaev, A.~S., \& Sherdonov, C.~T.\ 2002, \apj, 564, 1042 
\bibitem[Scherrer et al.(1995)]{1995SoPh..162..129S} Scherrer, P.~H., Bogart, R.~S., Bush, R.~I., et al.\ 1995, \solphys, 162, 129 
\bibitem[Sheeley et al.(1985)]{1985SoPh...98..219S} Sheeley, N.~R., Jr., Devore, C.~R., \& Boris, J.~P.\ 1985, \solphys, 98, 219 
\bibitem[Stenflo \& Kosovichev(2012)]{2012ApJ...745..129S} Stenflo, J.~O., \& Kosovichev, A.~G.\ 2012, \apj, 745, 129 
\bibitem[Tian et al.(2003)]{2003SoPh..215..281T} Tian, L., Liu, Y., \& Wang, H.\ 2003, \solphys, 215, 281 
\bibitem[Tlatov et al.(2010)]{2010ApJ...717..357T} Tlatov, A.~G., Vasil'eva, V.~V., \& Pevtsov, A.~A.\ 2010, \apj, 717, 357 
\bibitem[Ulrich et al.(1991)]{1991SoPh..135..211U} Ulrich, R.~K., Webster, L., Boyden, J.~E., Magnone, N., \& Bogart, R.~S.\ 1991, \solphys, 135, 211 
\bibitem[Ulrich et al.(2002)]{2002ApJS..139..259U} Ulrich, R.~K., Evans, S., Boyden, J.~E., \& Webster, L.\ 2002, \apjs, 139, 259 
\bibitem[Ulrich et al.(2009)]{2009SoPh..255...53U} Ulrich, R.~K., Bertello, L., Boyden, J.~E., \& Webster, L.\ 2009, \solphys, 255, 53 
\bibitem[Ulrich et al.(2010)]{2010SoPh..261...11U} Ulrich, R.~K., Parker, D., Bertello, L., \& Boyden, J.\ 2010, \solphys, 261, 11 
\bibitem[Wang \& Sheeley(1989)]{1989SoPh..124...81W} Wang, Y.-M., \& Sheeley, N.~R., Jr.\ 1989, \solphys, 124, 81 
\bibitem[Wang \& Sheeley(1991)]{1991ApJ...375..761W} Wang, Y.-M., \& Sheeley, N.~R., Jr.\ 1991, \apj, 375, 761 
\bibitem[Weber et al.(2011)]{2011ApJ...741...11W} Weber, M.~A., Fan, Y., \& Miesch, M.~S.\ 2011, \apj, 741, 11 
\end{thebibliography}
\end{document}